# Machine learning determines the Mg$_2$SiO$_4$ P–T phase diagram

Siyu Zhou[1,#], Daohong Liu[2#], Chuanyu Zhang[1,*], Yu He[2,3**],
Xuben Wang[4, ***], Xiaopan Zuo[1]

[1]College of Physics, Chengdu University of Technology, Chengdu, 610059, China

[2]State Key Laboratory of Critical Mineral Research and Exploration, Institute of Geochemistry, Chinese Academy of Sciences, Guiyang 550081, China.

[3]Center for High Pressure Science and Technology Advanced Research, Shanghai 201203, China

[4]College of Geophysics, Chengdu University of Technology, Chengdu, 610059, China

\* Corresponding author

\*\* Second Corresponding author

\*\*\* Third Corresponding author

\# These authors contribute equally

E-mail:(C. Zhang) zhangchuanyu10@cdut.edu.cn ; (Y. He)heyu@mail.gyig.ac.cn ;
(X. Wang)wxb@cdut.edu.cn

**Key Points:**

1. Based on first-principles GGA functionals, an initial dataset was constructed, the deep potential (DP) model and the neuroevolution potential (NEP) model suitable for both solid and liquid phases of Mg$_2$SiO$_4$ were developed using active learning and machine learning methods.

2. The phase diagram of the magnesium Mg$_2$SiO$_4$ system was determined using non-equilibrium thermodynamic integration and two-phase coexistence methods.

**Abstract**

Phase transitions among Mg$_2$SiO$_4$ and its high-pressure polymorphs (wadsleyite and ringwoodite) are central to mantle dynamics and deep-mantle material cycling. However, the locations and Pressure-Temperature (P-T) dependences of these phase boundaries remain debated, largely due to experimental limitations at extreme conditions and the high computational cost of first-principles free-energy calculations. Here, a machine-learning-potential driven workflow combining non-equilibrium thermodynamic integration (NETI) and two-phase coexistence simulations is employed to enable large-scale, long-timescale molecular dynamics sampling. Within this workflow, the melting curve of forsterite is evaluated and a complete P–T phase diagram is constructed. Relative to conventional ab initio approaches, this strategy reduces computational expense while retaining thermodynamic consistency in phase-stability assessment. The workflow is applicable to efficient evaluation of phase stability and thermodynamic properties in deep-Earth silicate systems.

**Plain Language Summary**

$Mg_2SiO_4$ is one of the most abundant minerals in the Earth's mantle, and its phase transitions to the high-pressure polymorphs wadsleyite and ringwoodite are closely related to key seismic discontinuities at depths of approximately 400–700 km below the Earth's surface. Thermodynamic constraints on these phase transitions are crucial for understanding the mantle's temperature structure, chemical composition, and deep-seated dynamic processes. This paper trains and constructs a machine learning interatomic potential that simultaneously describes the behavior of $Mg_2SiO_4$ solid and melt, and then performs large-scale molecular dynamics simulations to systematically reconstruct its phase diagram at a cost significantly lower than first-principles free energy calculations. The results show that the machine learning potential significantly improves computational efficiency while maintaining near-first-principles accuracy, providing an efficient and scalable computational pathway for studying the phase stability and melting behavior of deep-earth silicates.

# 1 Introduction

Silicate minerals are important components of the Earth's mantle. Among numerous silicate systems, $Mg_2SiO_4$ and its high-pressure polymorphs (wadsleyite and ringwoodite) are highly representative, which are the main mineral phases from the upper mantle to the mantle transition zone and are closely related to the mineralogical origin of the seismic discontinuities at 410 km and 520 km (Akaogi et al., 1989; Yu et al., 2008; H. Zhang et al., 2022). Therefore, a complete phase diagram that jointly constrains solid–solid and solid–liquid boundaries is essential to connect mineral physics with seismic observations and to evaluate how mantle temperature and composition modulate these discontinuities.

However, under the extreme temperature and pressure conditions associated with the Earth's mantle, experimental methods still face significant challenges in providing rigorous thermodynamic constraints for these crucial phase transitions. On the one hand, due to the limitations of experimental equipment and irreversible effects in phase transition kinetics, the system often fails to reach thermodynamic equilibrium. On the other hand, pivotal thermodynamic quantities such as entropy and Gibbs free energy cannot be directly measured experimentally and usually rely on indirect inference methods. Existing models of the $Mg_2SiO_4$ phase diagram largely depend on limited experimental data and specific thermodynamic models (Akaogi et al., 1989; Chopelas, 1991; Stixrude & Lithgow-Bertelloni, 2005), which inevitably introduces considerable uncertainty in determining the location of phase boundaries.

In contrast, first-principles calculations have been widely used to estimate free energy and phase stability under high-temperature and high-pressure conditions, and can provide preliminary predictions of phase boundary locations. However, in the high-temperature range, due to the gradually decreasing applicability of quasi-harmonic approximation (QHA) and the enhanced anharmonic effects of thermal vibrations, the phase boundary pressure still has a non-negligible uncertainty (Tsuchiya et al., 2004; Yu et al., 2007). At the same time, different exchange-correlation functionals and thermodynamic paths, such as QHA and the quasi-harmonic Debye model, show significant discrepancies in predicting the α-β phase boundary location and its Clapeyron slope (Liu et al., 2008; Yu et al., 2008). Furthermore, the computational cost of first-principles methods typically limits the feasible simulation scale to relatively small systems and short time scales, thus restricting sufficient sampling of complex phase space and thermal fluctuations.

Recent advances in machine learning potentials (MPLs) have opened new avenues for addressing these challenges, enabling the simulation of phase transitions and melting behavior under extreme temperature and pressure conditions (He et al., 2026.; Smith et al., 2021; L. Zhang et al., 2021), and demonstrating the feasibility of large-scale, long-time-scale dynamical simulations with near-first-principles accuracy (Behler, 2016). In particular, the MLPs such as deep potential (DP) model (L. Zhang et al., 2018), neuroevolution potential (NEP) model (Fan et al., 2021), and Gaussian approximation potentials (GAP) model (Bartók et al., 2010) have greatly expanded the phase spaceand temporal scales, making quantum level sampling of complex mineral systems possible at significantly reduced computational cost.

Based on these methodological advances, this study developed a high accuracy MLP for the $Mg_2SiO_4$ system, which was trained by employing high accuracy first-principles datasets. By combining this MLP with non-equilibrium thermodynamic integration (NETI) free energy calculations. Therefore, the deep potential molecular dynamics (DPMD) simulations were performed to calculate the phase transition points and melting points of the α, β, and γ phases, and to construct the Pressure-Temperature (P–T) phase diagram of $Mg_2SiO_4$ under mantle transition zone conditions. This study provides new quantitative constraints on the phase stability and Clapeyron slope of $Mg_2SiO_4$ polymorphs.

## 2. Computational simulation details

### 2.1 First-principles dataset preparation

The initial training dataset was constructed using the deep potential generator DP-GEN package (Y. Zhang et al., 2020) workflow integrated with ABACUS (3.9.0.2), where structural perturbations and short-time molecular dynamics simulations were employed to sample representative configurations. All first-principles calculations were performed using the linear combination of atomic orbitals (LCAO) basis set (M. Chen et al., 2010; P. Li et al., 2016), SG15 optimized norm-conserving Vanderbilt (ONCV) pseudopotentials (Hamann, 2013), and the Perdew–Burke–Ernzerhof (PBE) exchange–correlation functional (Perdew et al., 1996). A plane-wave energy cutoff of 100 Ry was adopted, and the self-consistent field (SCF) convergence criterion for the charge density was set to $1 \times 10^{-6}$ Ry. The k-point grid spacing was chosen to be 0.2 Å$^{-1}$, and a time step of 1 fs was used in all MD simulations. Supercells of the α, β, and γ phases contained 112, 112, and 56 atoms, respectively. The initial dataset was subsequently expanded through 12 iterations of active learning.

In addition, ABACUS was employed to enhance liquid state sampling by performing supplementary NPT ensemble ab initio molecular dynamics (AIMD) simulations at 10 GPa and 5000 K, from which representative liquid configurations were extracted. The datasets generated from active learning and enhanced sampling were filtered using the NepTrainKit package (C. Chen et al., 2025), resulting in a curated dataset comprising 2516 configurations. This dataset was further refined through four additional iterative training cycles, extending the explored thermodynamic space to pressures up to 35 GPa and temperatures up to 5000 K. The final dataset was then used for long-time-scale training of the machine-learning interatomic potentials.

### 2.2 Machine Learning

The collected dataset was subsequently used for long-time training with the DeePMD-kit package (Wang et al., 2018; Zeng et al., 2023, 2025; D. Zhang et al., 2024). The *se_atten_v2* descriptor was adopted, with the *rcut_smth* set to 0.5 Å and the interaction cutoff radius extended to 7.5 Å to better capture longer-range interactions. The embedding network was figured with layer sizes of {25, 50, 100}, while the fitting neural network consisted of three hidden layers with 240 neurons each ({240, 240, 240}). The total number of training steps was set to 5,000,000 to ensure robust convergence of the potential. In parallel, the initial dataset was subjected to an energy shifting procedure, and the neuroevolution potential (NEP) models were trained using the same configurations.

## 2.3 Deep potential molecular dynamics simulation

The large scale molecular dynamics simulator LAMMPS package (Plimpton, 1995) was coupled with the DP model for calculations. The α phase used a system of 7056 atoms, the β phase 7840 atoms, and the γ phase 7000 atoms. The system temperature and pressure were controlled using the Langevin and Nosé–Hoover thermostat (Hoover, 1985; Nosé, 2002; Schneider & Stoll, 1978), respectively. First, approximately 20 ps of pre-equilibration was performed under the NVT ensemble to stabilize the system temperature and prevent rapid expansion of the initial lattice due to pressure coupling. Subsequently, approximately 100 ps of simulation was performed under NPT conditions to obtain equilibrium lattice parameters at specific temperatures and pressures. After the system volume and structure had fully converged, approximately 100 ps of time averaging of atomic positions was performed to obtain a time-averaged equilibrium configuration, thereby optimizing the crystal structure. Subsequently, an additional 100-picosecond sampling phase was used to calculate the mean square displacement (MSD) of Mg, Si, and O atoms. This MSD was used to characterize the atomic vibration characteristics and derive the corresponding Einstein crystal spring constant $k$, and to provide the initial configuration for subsequent calculations using the non-equilibrium thermodynamic integration (NETI) free energy method (Cajahuaringa & Antonelli, 2022; Freitas et al., 2016; Paula Leite & De Koning, 2019) . The $k$ is defined as:

$$k = \frac{3k_\mathrm{B}T}{\langle(\Delta \mathbf{r})^2\rangle} \qquad (1)$$

where $k_\mathrm{B}$ is the Boltzmann constant, $T$ denotes the system temperature, and $\langle(\Delta r)^2\rangle$ represents the mean square displacement (MSD) of atoms relative to their time-averaged equilibrium positions. The angular brackets indicate statistical averaging over time and over all atoms in the system.

The melting points was determined using the two-phase coexistence (TPC) method (Morris et al., 1994). All molecular dynamics simulations were based on a system of 3360 atoms in the α and β phases. For each target pressure, the system was first equilibrated in the NPT ensemble for 10 ps at an initial temperature close to the expected melting point, allowing sufficient relaxation of the cell volume and removal of residual stresses. Subsequently, the supercell was divided into solid and liquid regions along the z direction, and region-specific temperature control was applied in the NVT ensemble for 25 ps to construct an initial solid–liquid interface. At this stage, the temperature of the liquid phase region was set to 5000 K to ensure complete melting of the liquid phase under superheated conditions. Subsequently, the temperature of the entire system was adjusted to 1.1 melting temperature, and the simulation was continued for an additional 15 ps to obtain a structurally stable two-phase configuration with a well-defined solid–liquid interface. After a stable two-phase configuration was established, the simulation was switched to the NPH ensemble with anisotropic pressure control, allowing the system to evolve freely under the target pressure. This production run was performed for 250 ps to adequately sample the long-time dynamical behavior of the solid–liquid interface. The melting temperature was determined as the average system temperature over the final 50 ps after convergence was achieved. To reduce statistical fluctuations and improve the reliability of the melting-point determination, the entire TPC procedure was independently repeated five times at each pressure using different random seeds, and the average value was taken as the final melting temperature.

## 3. Results

### 3.1. Benchmark test of MLP models

The accuracy of the DP model was assessed against DFT reference data in terms of energies and forces (Figure 1). The energy errors for each phase were constrained to 1.11–5.92 meV·atom$^{-1}$ (Figure S1). The performance of the NEP model, which was subsequently used as an auxiliary model in later analyses, is reported in Figure S2. For both MLP models, deviations from the DFT references remain consistently low, indicating that the underlying potential energy surface is reproduced with high fidelity across the sampled configuration space.

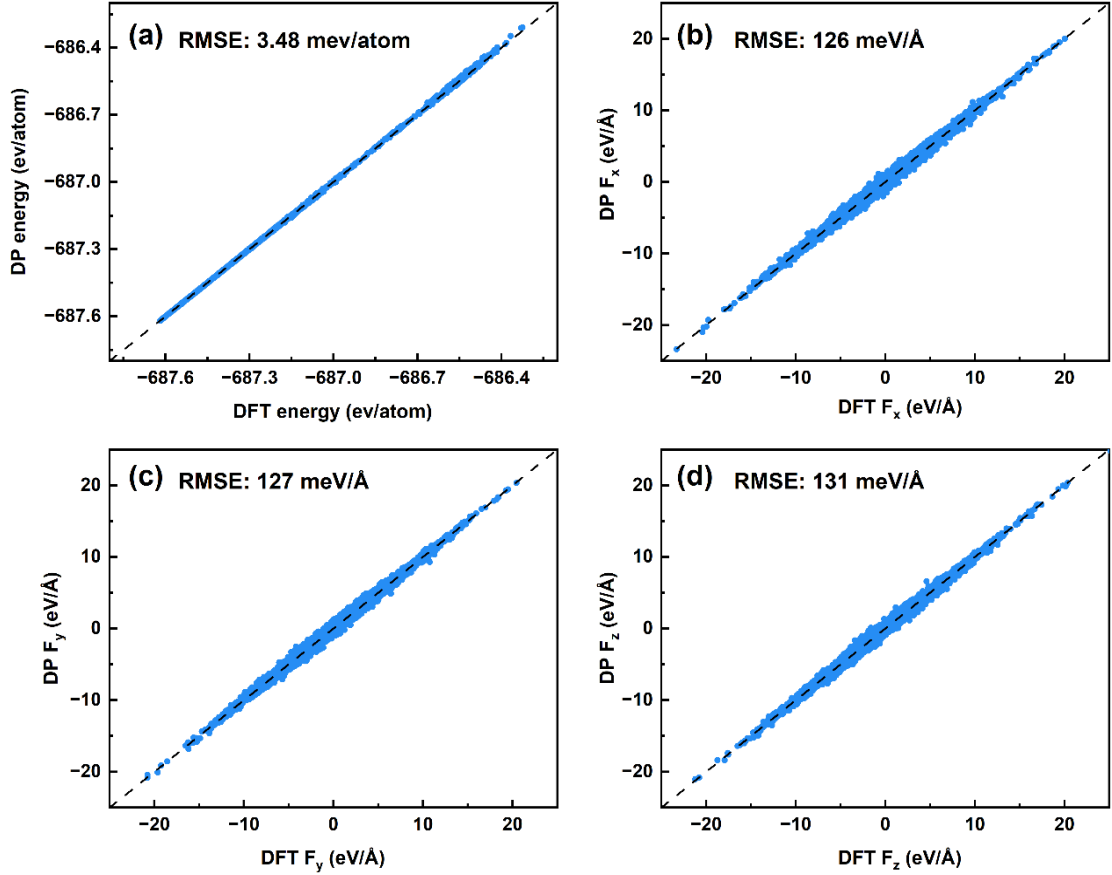

**Figure 1. Comparison of energies and forces predicted by the DP potential with DFT reference data.** (a) potential energy per atom and (b–d) force components in the x, y, and z directions. The dashed black line indicates the ideal agreement y=x, demonstrating excellent consistency between DP predictions and DFT calculations across the entire dataset.

Beyond matching DFT energies and forces, it is also necessary to verify whether physically relevant structural descriptors are accurately reproduced. To this end, radial distribution functions (RDF) were computed for the α, β, γ, and liquid phases using both DP and NEP, and were compared directly with DFT results (Figure 2). Close agreement is observed across all phases, suggesting that short-range order dominated by Mg–O and Si–O bonding is well captured by the machine-learning models. The RDF is defined as,

$$g(r) = \frac{dN}{4\pi r^2 dr} / (\rho dV) \qquad (2)$$

where $dN$ is the number of target atoms within a spherical shell of thickness $dr$ at distance $r$, $\rho$ is the system number density, and $dV$ is the shell volume.

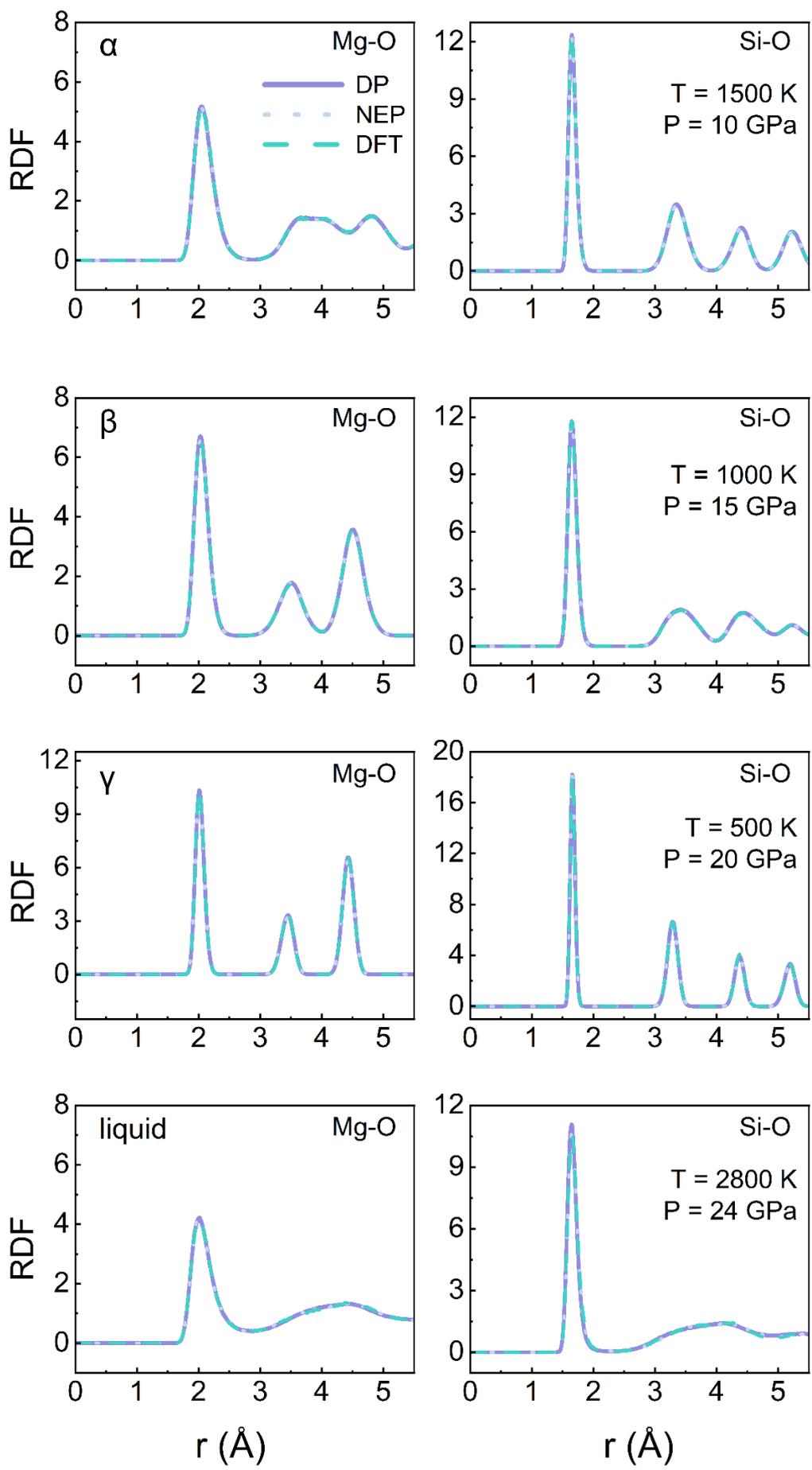

**Figure 2. Comparison of radial distribution functions (RDFs) for different phases of Mg₂SiO₄ calculated using various potentials.** The panels illustrate the Mg-O and Si-O pair distributions for the α phase (1500 K, 10 GPa), β phase (1000 K, 15 GPa), γ phase (500 K, 20 GPa), and liquid phase (2800 K, 24 GPa). The solid line (DP), dotted line (NEP), and dashed line (DFT) represent the results obtained from the Deep Potential, Neuroevolution Potential, and Density Functional Theory, respectively.

Consistency in thermodynamic properties was further evaluated through density predictions. Densities of the three solid polymorphs were first calculated at 500 K under pressures of 5, 10, and 20 GPa using the DP and DFT methods (Figure S3), and good agreement was obtained. The assessment was then extended to a broader domain of 500–2500 K and 0–40 GPa, where equilibrium densities were predicted using both DP and NEP. Based on these results, equations of state (EOS) for the three phases were fitted using the third-order Birch–Murnaghan form. As shown in Figure 3, agreement between DP and NEP remains robust even at elevated temperatures and pressures, supporting their applicability in the high P–T stability fields examined in this work.

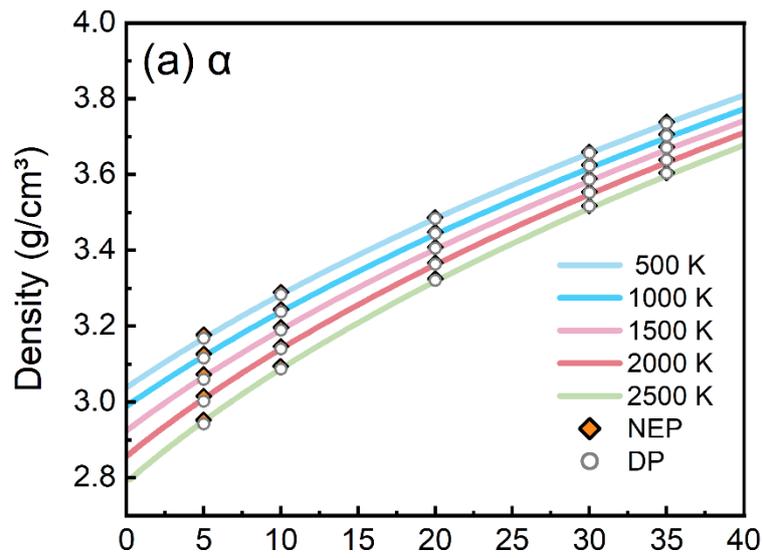
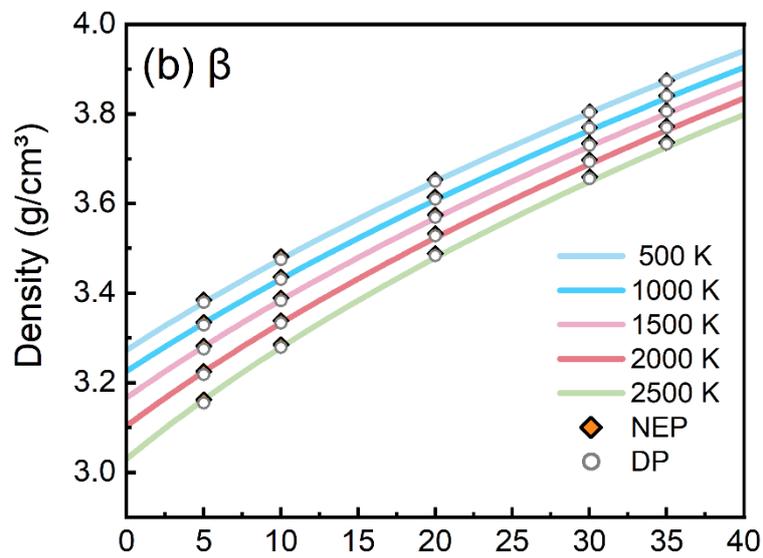
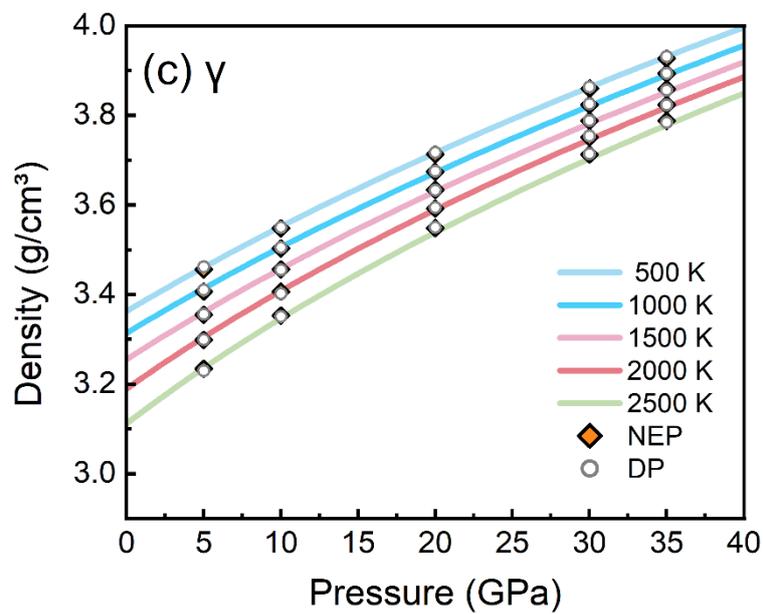

**Figure 3. Pressure dependence of density for the three solid polymorphs of Mg₂SiO₄ at different temperatures**. (a) α phase, (b) β phase, and (c) γ phase. Solid lines represent densities obtained from fits to the third-order Birch–Murnaghan equation of state, with different colors corresponding to 500 K (light blue), 1000 K (blue), 1500 K (pink), 2000 K (red), and 2500 K (green). Symbols denote equilibrium densities calculated using machine-learning interatomic potentials, with diamonds indicating NEP results and open circles indicating DP results.

### 3.2. Clapeyron slopes and melting curve parameters

Before determining the solid phase-transition points, the convergence of the NETI was estimated by employing different switching time (Figures S4–S5). Subsequently, an extended adiabatic switching (AS) approach (Freitas et al., 2016) was applied to determine the transition pressures over a range of temperatures by evaluating the pressure dependence of the Gibbs free energy differences (Figures S6 – S8).

The resulting Clausius-Clapeyron slopes indicate that the α–β phase transition slopes predicted by both the DP and NEP methods are greater than the estimated value obtained using the GGA‐PBE combined with QHA first-principles calculation method (Yu et al., 2008), whereas the β–γ slopes are smaller. Specifically, the α–β slope was determined to be 2.81 (2.82) MPa/K for DP and NEP, while the β–γ slope was predicted to be 2.84 (2.00) MPa/K for (DP) and (NEP). Despite these differences relative to GGA–PBE combined with QHA, the predicted α–β slopes remain within the range of available experimental constraints, whereas the β–γ slopes tend to be slightly lower than reported values (Akaogi et al., 1989; Inoue et al., 2006; Katsura & Ito, 1989a; Suzuki et al., 2000). Overall, in the high-pressure and high-temperature regime, phase boundaries predicted by DP and NEP exhibit consistent transition trends.

For the melting curve, one of the equation proposed by Kechin was adopted (Kechin, 2001). one form of which is as follows:

$$T_m(p) = T_0 \left(1 + \frac{p - p_0}{A}\right)^B \exp\left(-\frac{p - p_0}{C}\right) \qquad (3)$$

where $T_0$ and $P_0$ denote the reference melting temperature and pressure, and $a$, $b$, and $c$ are fitting parameters. The fitted parameters based on the DP (NEP) model are $a = 18.31(21.17)$ GPa, $b = 0.42(0.76)$, and $c = 78.07(44.61)$ GPa.

### 4. Discussion

#### 4.1. Forsterite melting curve

The melting curve of forsterite ($\alpha$ phase) is presented in Figure 4, fitted using melting points determined from the two-phase coexistence simulations (Figure S9). The predicted melting point at atmospheric pressure is slightly lower than the value measured experimentally by Bowen, et al (1914). In contrast, the predictions in the 3–5 GPa range are found to be in good agreement with the measurements of Davis, et al (1964).

Overall, melting temperatures obtained with the NEP model are systematically lower than those predicted by the DP model. It should be noted, however, that neither machine-learning model captures the feature reported by De Koker et al. (2008), where a Clausius–Clapeyron integration analysis combined with the LDA functional suggested that the melting temperature reaches a maximum at 13 GPa and 2550 K, followed by a negative slope at higher pressures.

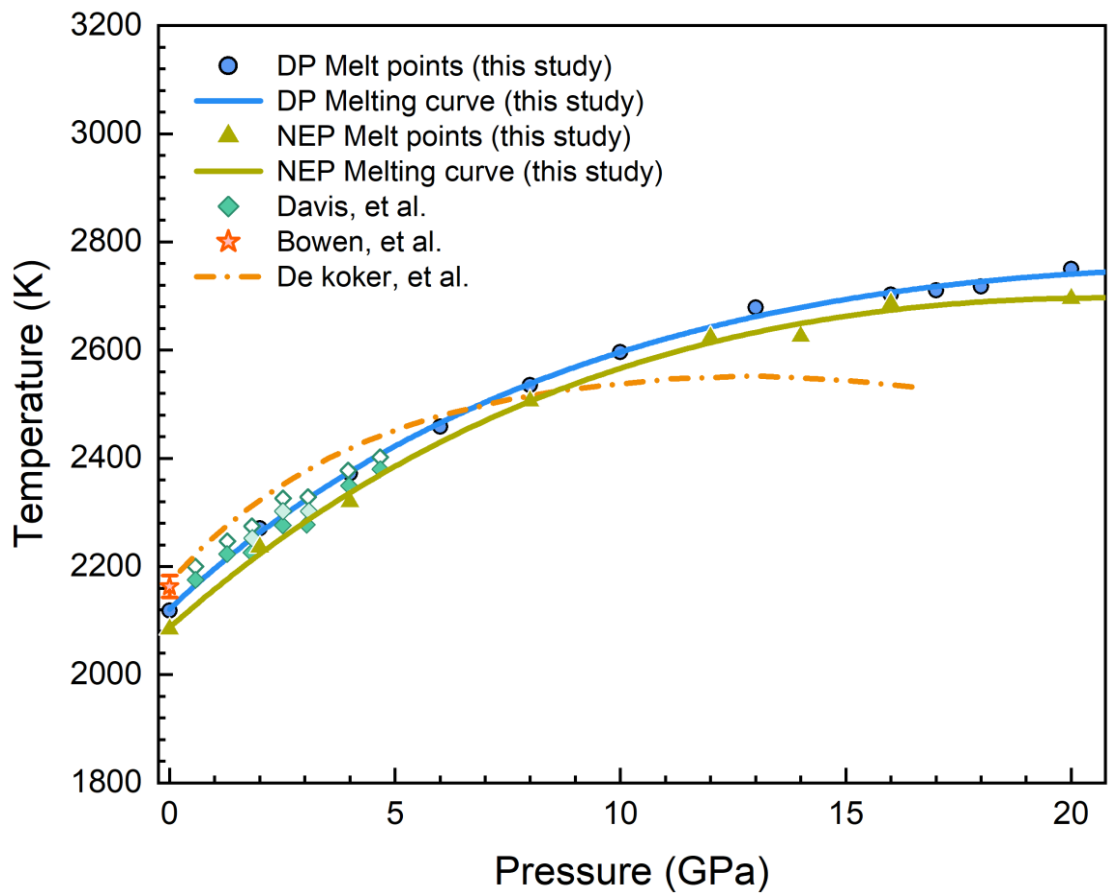

**Figure 4. Melting curve $T_m(P)$ of forsterite ($\alpha$ phase)** Blue circles denote melting points obtained in this study using the DP model, and the blue solid line shows the corresponding fitted melting curve. Olive triangles denote melting points obtained using the NEP model, and the olive solid line shows the corresponding fitted melting curve. Previous experimental data are shown as green diamonds (Davis et al., 1964), orange short dashed line (De Koker et al., 2008), and orange star (Bowen et al., 1914).

**4.2. Phase diagram**

The phase diagram constructed from the DP and NEP models under the methodological framework adopted in this study is shown in Figure 5(a). When compared with previous computational results, the discrepancies in the solid–solid boundaries are primarily associated with the α - β boundary. Specifically, the transition points predicted by the DP model are shifted toward higher pressures relative to the DFT combined with the QHA results. This offset may partly be attributed to systematic finite-size effects arising from differences in the simulation cell size. More importantly, it is likely indicative of the limited capability of the QHA to adequately represent minerals that exhibit pronounced anharmonicity, such as forsterite (Gillet et al., 1991).

Notably, within the QHA breakdown regime (Wentzcovitch et al., 2004), the DP calculations predict higher α–β transition pressures than QHA (Figure S10(a)), whereas the predicted β–γ boundary pressures are lower than those from QHA (Figure S10(b)). The calculation results using the NEP model showed the same qualitative trends, providing additional support for this observation. Together, these results indicate that high-temperature QHA deviations in the $Mg_2SiO_4$ system cannot be captured by a uniform pressure-shift correction across different phase boundaries. Instead, the anharmonic contributions to the free energy can differ substantially between phases. Similar behaviors have been reported in polymorphic systems of MgO, $MgSiO_3$, and $SiO_2$ (G. Li et al., 2023; Soubiran & Militzer, 2020; Z. Zhang & Wentzcovitch, 2022), thereby producing distinct offset trends for phase boundaries across different stability fields.

Overall, the forsterite phase diagram obtained with the machine-learning models is broadly consistent with the experimentally extrapolated phase diagram (Figure 5(b)) in terms of both the transition trends and the overall topological structure, which supports the predictive reliability of machine-learning models for complex silicate systems. Relative to experimental constraints, the systematic shift of the predicted boundaries toward higher pressures is attributed to intrinsic limitations of the underlying density-functional approximation used to generate the training reference data, rather than to distortions introduced during potential fitting. A similar systematic drift in phase boundaries associated with the accuracy limits of the reference functional has also been documented in deep potential studies of the water phase diagram (L. Zhang et al., 2021).

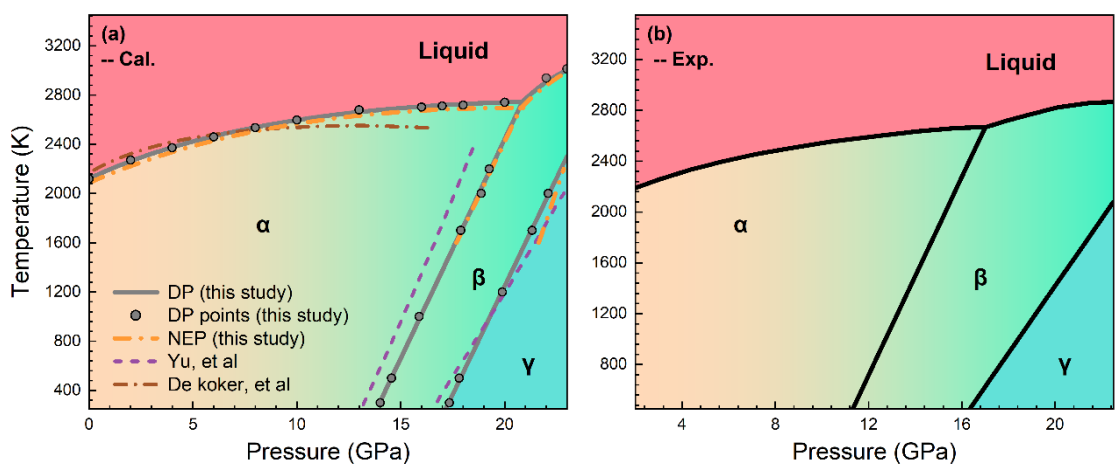

**Figure 5. P-T phase diagram of $Mg_2SiO_4$ comparing calculated and experimental phase boundaries. (a)** Calculated P–T phase diagram. Solid grey lines denote phase boundaries predicted using the DP potential, and grey filled circles mark the phase-transition points determined from Gibbs free-energy crossings. NEP predictions are shown as yellow dash–dot lines. DFT (GGA) combined with QHA is shown as a purple dashed line (Yu et al., 2008). The melting curve obtained using the LDA functional and the Clausius–Clapeyron integration method is shown as a brownish red short dashed line (De Koker et al., 2008). **(b)** Experimentally extrapolated phase diagram compiled

from previous studies (Andrault & Bonhifd, 1995; Bouibes & Zaoui, 2020; Guyot & Reynard, 1992; Katsura & Ito, 1989b; Rouquette et al., 2008; Tokár et al., 2013; Horiuchi et al., 1981).

## 5. Conclusion

This study employed machine learning combined with non-equilibrium free energy and two-phase methods, using larger systems and longer simulation times, to derive the forsterite melting curve and construct a complete phase diagram of forsterite. Both the DP and NEP models showed consistent predictive results regarding the overall phase transition trend. This further supports the reliability of this method for rapidly and comprehensively determining the complete phase diagrams of complex minerals under high-temperature and high-pressure conditions. In the future, the machine learning models obtained using this method will be extended to iron-bearing olivine and more lower mantle mineral phases, and combined with non-equilibrium molecular dynamics (NEMD) methods to better study thermal transport properties. This method will also be extended to research in related planetary science fields.

## Data Availability

The data underlying this paper will be shared on reasonable request to the corresponding author.

## Declaration of Interest Statement

No conflict of interest exits in the submission of this manuscript, and manuscript is approved by all authors for publication. I would like to declare on behalf of my co-authors that the work escribed was original research that has not been published previously, and not under consideration for publication elsewhere, in whole or in part.

## Supplementary Information

See attached file.

## Acknowledgment

This work is supported by the National Natural Science Foundation of China (42230311), and the Sichuan Students' innovation and entrepreneurship training program (S202510616025). All the ab initio simulations were performed on Chengdu Supercomputing Center.